\documentclass[a4paper,fleqn,usenatbib]{mn2e}

\usepackage[T1]{fontenc}
\usepackage{ae,aecompl}

\usepackage{graphicx}	
\usepackage{amsmath}	
\usepackage{amssymb}	

\def\la{\;
\raise0.3ex\hbox{$<$\kern-0.75em\raise-1.1ex\hbox{$\sim$}}\; }
\def\ga{\;
\raise0.3ex\hbox{$>$\kern-0.75em\raise-1.1ex\hbox{$\sim$}}\; }

\newcommand{\daa}{$\Delta\alpha/\alpha$}
\newcommand{\kms}{km~s$^{-1}$}
\newcommand{\etal}{{et al.}}

\title[FST method to probe $\alpha$ at high redshifts]{Fine-structure transitions as a tool for 
studying variation of $\alpha$ at high redshifts}

\author[S. A. Levshakov \& M. G. Kozlov]{
S. A. Levshakov,$^{1,3,4}$\thanks{E-mail: lev@astro.ioffe.ru}
M. G. Kozlov$^{2,3}$
\\
$^{1}$Ioffe Physical-Technical Institute, 194021 St.~Petersburg, Russia\\
$^{2}$Petersburg Nuclear Physics Institute, 188300 Gatchina, Russia\\
$^{3}$St.~Petersburg Electrotechnical University ``LETI'', 197376 St.~Petersburg, Russia\\
$^{4}$ITMO University, 191002 St.~Petersburg, Russia
}

\date{Accepted XXX. Received YYY; in original form ZZZ}

\pubyear{2017}

\begin{document}
\label{firstpage}
\pagerange{\pageref{firstpage}--\pageref{lastpage}}
\maketitle

\begin{abstract}
Star-forming galaxies at high redshifts  
are the ideal targets to probe the hypothetical variation of the
fine-structure constant $\alpha$ over cosmological time scales.
We propose a modification of the alkali doublets method
which allows us to search for variation in $\alpha$ 
combining far infrared and submillimeter spectroscopic observations.
This variation manifests as velocity offsets between the observed 
positions of the fine-structure and gross-structure transitions
when compared to laboratory wavelengths.
Here we describe our method whose sensitivity limit to the fractional changes in
$\alpha$ is about $5\times10^{-7}$.
We also demonstrate that current spectral observations of hydrogen and [C\,{\sc ii}]
158 $\mu$m lines provide an upper limit on
$|\Delta\alpha/\alpha| \la 6\times10^{-5}$ at redshifts $z = 3.1$ and $z = 4.7$.
\end{abstract}


\section{Introduction}

The Einstein equivalence principle (EEP) postulates that fundamental
physical laws are invariant in space and time. 
However, some of new theories beyond the Standard Model (SM) of particle physics suggest
a violation of the EEP (for a review, see, e.g., Liberati 2013). 
In particular, a changing dimensionless fine-structure constant,
$\alpha = e^2/\hbar c$, accompanied by variation in other coupling
constants can be associated with a break of local Lorentz invariance
(Kosteleck\'y \etal\ 2003). 
Thus, experimental validation of EEP is extremely important allowing us 
both to test limits of the standard theory, i.e. the quantum electrodynamics, 
and to probe the applicability of the new theories beyond the SM. 
Astronomical measurements seem to be the most suitable tool for this purpose
thanks to their accuracy and the possibility to operate within and 
outside of our Galaxy and over time intervals comparable to the age of 
the Universe ($\sim 10^{10}$ yr). 

In astrophysics, the validity of EEP was tested 
through the observed time delays between different energy bands from blazar flares,
gamma-ray bursts, and fast radio bursts which constrain
the numerical coefficients of the parameterized post-Newtonian formalism, 
such as the parameter $\gamma$ 
accounting for how much space-curvature is produced by unit rest mass
(e.g., Wei \etal\ 2015; Petitjean \etal\ 2016).
It was shown that general relativity, which predicts $\gamma \equiv 1$,
is obeyed to the level of $\sim 10^{-8}$ (Gao \etal\ 2015; Wei \etal\ 2016).

More than two decades ago, Levshakov (1992) proposed a program of high resolution spectral
observations of extragalactic sources~-- quasars -- with a new
generation of giant telescopes aimed at differential measurements of
\daa \footnote{ $\Delta\alpha/\alpha = (\alpha_{\rm z} -
\alpha)/\alpha$, with $\alpha_{\rm z}$ being the fine-structure
constant at redshift $z$, and $\alpha$ is the present-day value.}
from alkali doublets. 
The alkali doublet method uses the ratio of the difference between fine-structure
lines to their average wavelength 
(Savedoff 1956; Bahcall \& Salpeter 1965; 
Bahcall \& Schmidt 1967; Bahcall \etal\ 1967;
Levshakov 1994; Bahcall \etal\ 2004). 
Later on, the alkali doublet method was supplemented
with the so-called ``many-multiplet'' method which utilizes absolute
laboratory wavelengths arising from 
many different multiplets in different ions (Dzuba, Flambaum, \& Webb 1999a,b).
The alkali doublet method was applied both to emission-line spectra of
distant galaxies and to absorption-line spectra of the intervening quasar absorbers.
The many-multiplet method dealt with quasar absorption-line spectra only.
This may lead spectral lines sampling quite a different interstellar medium
in the quasar absorption-line systems and distant galaxies seen in emission.

Since that time there appeared a number of controversial
results suggesting a time dependence of $\alpha$ and
restricting \daa\ values to the same level of a few $10^{-6}$
(see, e.g., Molaro \etal\ 2013, and references cited therein). 
All measurements were carried out on basis of optical
observations performed at the VLT, Keck, and Subaru telescopes.
Recently it became clear that the above mentioned inconsistencies were caused by
systematic effects in the calibration of the wavelength scale of quasar
spectra (Whitmore \& Murphy 2015; Evans \etal\ 2014). 
The tightest present-day upper limit on the relative variation in $\alpha$, obtained
from optical spectra of quasars, is limited to a few $10^{-6}$ at
$z \sim 2-3$ (e.g., Quast \etal\ 2004; Levshakov \etal\ 2006; Agafonova \etal\ 2011;
Bonifacio \etal\ 2014). 
A further improvement~-- approximately by an order of magnitude~-- can be
expected only with the new high-resolution-ultra-stable spectrograph
ESPRESSO at the VLT, whose commissioning will start in 2017 (Leite \etal\ 2016).

Together with astronomical studies the fundamental physical laws were
tested for the passed decade in numerous laboratory experiments with atomic clocks.
For this purpose, frequencies of two or more lines with different
dependence on the fundamental constants are compared to
each other in time to monitor a relative frequency shift, $\Delta \omega$.
Laboratory experiments provide extremely high sensitivity with
fractional uncertainties of $\Delta \omega/\omega \sim 10^{-16} - 10^{-17}$
(Tyumenev \etal\ 2016; Nemitz \etal\ 2016; Huang \etal\ 2016) 
and even at the $10^{-18}$ level 
(Nisbet-Jones \etal\ 2016; Huntemann \etal\ 2016; Nicholson \etal\ 2015).
For the time scale of the order of one year the fractional change in $\alpha$
was restricted at the 
$\dot{\alpha}/\alpha \sim 10^{-17}$ yr$^{-1}$ level
(Godun \etal\ 2014; Rosenband \etal\ 2008).

To the same category of terrestrial experiments belongs the Oklo phenomenon~--
the uranium mine at age $2\times10^9$ yr
in Gabon which provides the tightest terrestrial bound on $\alpha$
variation: $|\Delta\alpha/\alpha| < 1.1\times10^{-8}$,
or $\dot{\alpha}/\alpha < 5\times10^{-18}$ yr$^{-1}$ (Davis \& Hamdan 2015).
 
An independent and important complement to laboratory experiments with atomic
clocks and optical observations of
quasars provide far infrared (FIR) and sub-mm observations of distant galaxies. 
Advantages of micro-wave observations are the absence of 
systematic errors in calibrating the wavelength scale 
and in a higher sensitivity of atomic transitions to changes in
$\alpha$ in this spectral range (Kozlov \etal\ 2008). 
For instance, in the case of the $^2P_J$ multiplet of C\,{\sc ii} there is 
a $J = 3/2 \rightarrow 1/2$ fine-structure (FS) transition (158 $\mu$m)
which is about 30 times more sensitive to changes in $\alpha$ than
UV lines of atoms and ions employed in optical spectroscopy. 

The C\,{\sc ii} ion is one of the main cooling agents of the interstellar
medium (ISM) and its above mentioned transition forms the brightest
line detected up to very high cosmological redshifts $z \sim 3-8$
(e.g., Carniani \etal\ 2017; Umehata \etal\ 2017; Knudsen \etal\
2017; Aravena \etal\ 2016; Pavesi \etal\ 2016; Knudsen \etal\ 2016;
Maiolino \etal\ 2015; Capak \etal\ 2015; Willott \etal\ 2015;
Riechers \etal\ 2014; Carilli \etal\ 2013; Venemans \etal\ 2012; Maiolino \etal\ 2005). 
With the ALMA (Atacama Large Millimeter/submillimeter Array) 
facility one might expect that observations of C\,{\sc ii} will become possible
even for galaxies at extreme redshifts $z \sim 20$ (de Blok \etal\ 2016). 
Thus, FS transitions advance to an important tool in probing
the fundamental physics at high redshifts.

The FS lines of [C\,{\sc ii}] 158 $\mu$m and [C\,{\sc i}] 370 $\mu$m
were used in conjunction with the pure rotational CO transitions
to put constraints on the ratio
$F = \alpha^2/\mu$ (where $\mu$ is the electron-to-proton mass ratio)
at $z = 6.42$ and 4.69 (Levshakov \etal\ 2008), $z = 5.2$
(Levshakov \etal\ 2012), and $z = 2.79$ (Wei\ss\ \etal\ 2012).
In these measurements, the most stringent upper limit on $F$ 
was obtained by Wei\ss\ \etal: $|\Delta F/F| < 10^{-5}$.

In the present letter, 
we discuss another possibility of using the atomic FS transitions~-- to probe
the fractional changes not in a combination of constants, but directly in $\alpha$.
The proposed ``FST method'' (Fine Structure Transition) is based on the measurements of
the radial velocity offset, $\Delta V = V_{\rm opt} - V_{\rm fs}$,
between optical and FIR fine-structure lines within ground multiplets.

The rest of this letter is arranged as follows.
In Section 2, we briefly describe the FST method and show its applicability
to recently published spectral observations of a dusty star-forming galaxy
observed at $z = 3.1$, and a group of galaxies at $z = 4.7$. 
In Section 3, we summarize our results.

\section{FST method and results}

We consider FS transitions in the ground state multiplet of atoms
and/or ions and illustrate results for the ${}^2P_J$ multiplet of
the C\,{\sc ii} ion. 
In this case there is only one FS line ${}^2P_{3/2}
\rightarrow {}^2P_{1/2}$ at wavelength $\lambda = 157.7409(1)$ $\mu$m
(the corresponding frequency 1900.5369(13) GHz is given in  Cooksy \etal\ 1986). 
For reference we use the atomic hydrogen H$_\beta$
transition ($n=4 \rightarrow n=2$) at vacuum wavelength 
$\lambda = 4862.721(1)$ \AA, which is recommended in the Sloan Digital Sky Survey\footnote{
http://classic.sdss.org/dr7/products/spectra/vacwavelength.html}.

[C\,{\sc ii}] emission arises predominantly from the warm neutral
regions between the dense molecular clouds and H\,{\sc ii} regions and, thus,
its distribution is confined to the optical disks of galaxies.
Observations with high spatial resolution of H\,{\sc ii} regions in the nearby galaxy M33
show that the radial velocities of the [C\,{\sc ii}] and H\,{\sc i} 
emitting clouds appear to match each other with an offset of a few km~s$^{-1}$
(e.g., Braine \etal\ 2012; Mookerjea \etal\ 2016). 
From this, we might expect that global velocity profiles 
of hydrogen Balmer and [C\,{\sc ii}] lines 
integrated over whole disks of high-redshift galaxies tend
to match each other as well. 
However, any other atomic transitions, which trace the spatial
distribution of C\,{\sc ii}, can be utilized as well.
We note that numbers in parentheses in the wavelength values correspond
to an error at the last digit. Converted into velocity scale these
errors are equal to $\sigma_{\rm v}$(C\,{\sc ii}) = 0.19 \kms\ and
$\sigma_{\rm v}$(H$_\beta$) = 0.06 \kms.

The non-relativistic atomic energy differences (the gross
structure) and the relativistic corrections due to the spin-orbit
interaction (the fine-structure) are proportional to $\alpha^2$ and
$\alpha^4$, respectively (e.g., Sobelman 1979). 
This means that the
ratio of the frequencies of the fine-structure splitting to the
optical transition is proportional to the fine-structure constant squared.

If $\omega_{\rm fs,z}$ is the frequency of the fine-structure
splitting, and $\omega_{\beta,{\rm z}}$ is the frequency of the
hydrogen H$_\beta$ line, and $\alpha_{\rm z}$ is the value of the
fine-structure constant at redshift $z$, then
\begin{equation}
\frac{\omega_{\rm fs,z} \omega_{\beta}}{\omega_{\beta,{\rm z}} \omega_{\rm fs}} =
\left( \frac{\alpha_{\rm z}}{\alpha}\right)^2
\approx 1 + 2\frac{\Delta\alpha}{\alpha}\, ,
\label{Eq1}
\end{equation}
where $\omega_{\rm fs}$, $\omega_{\beta}$, and $\alpha$ are the
present-day values, and $\Delta\alpha = \alpha_{\rm z} - \alpha$.
Here we assume that $|\Delta\alpha/\alpha| \ll 1$.

If $\Delta\alpha \neq 0$ and $\omega'_{\rm fs}$, $\omega'_{\beta}$
are the observed frequencies from a distant object, then the
apparent redshifts of the FS and H$_\beta$ lines are given by
\begin{equation}
1+z_1 = \omega_{\rm fs}/\omega'_{\rm fs}\, ,
\label{Eq2}
\end{equation}
and
\begin{equation}
1+z_2 = \omega_\beta /\omega'_\beta\, .
\label{Eq3}
\end{equation}
Substituting equations (\ref{Eq2}) and (\ref{Eq3}) into equation
(\ref{Eq1}) and carrying out the straightforward algebra, we find
\begin{equation}
\frac{\Delta\alpha}{\alpha} = \frac{\Delta z}{2(1+\bar z)} \equiv \frac{\Delta V}{2c}\, ,
\label{Eq4}
\end{equation}
where $\bar{z}$ is the mean redshift, $\Delta z = z_2 - z_1$,
$c$ is the speed of light, and $\Delta V= V_2-V_1$ is the
velocity offset. Here we assume that $|\Delta z| \ll 1$.

Equation (\ref{Eq4}) shows that
the key point in probing \daa\ is the accurate determination of the
radial velocity offset between emission lines of two
elements which should trace closely each other.
If both line centers are measured
with the same uncertainty $\sigma_{\rm v}$, then the error of
the offset $\Delta V$ is
$\sigma_{\scriptscriptstyle \Delta V} = \sqrt{2}\sigma_{\rm v}$, which gives
the error $\sigma_{\scriptscriptstyle \Delta\alpha/\alpha}$ of the fractional change in $\alpha$:
\begin{equation}
\sigma_{\scriptscriptstyle \Delta\alpha/\alpha} = \sigma_{\rm v}/(\sqrt{2}c)\, .
\label{Eq5}
\end{equation}

If the quoted above accuracy level for these lines ($\sigma_{\rm v} < 0.2$ \kms)
is the only source of uncertainties in the line centers,
then the limiting sensitivity of the FST method for the combination of the 
[C\,{\sc ii}] 158 $\mu$m and H$_\beta$ lines
is $\sigma_{\scriptscriptstyle \Delta\alpha/\alpha} \la 5\times10^{-7}$.
Of course, in real observations, the measurement error is larger.

To illustrate the FST method, we consider a limit on \daa\
towards a dusty star-forming galaxy embedded in a giant Ly-$\alpha$ blob (LAB)
at $z = 3.1$ where strong [C\,{\sc ii}] 158 $\mu$m and H$_\beta$ emission lines were recently
detected by Umehata \etal\ (2017), and Kubo \etal\ (2015), respectively.
This galaxy, called LAB1-ALMA3, was discovered in
the ALMA dust continuum survey at 850 $\mu$m (Geach \etal\ 2016).

According to Umehata \etal\ (2017),
the velocity offset between the [C\,{\sc ii}] 158 $\mu$m and H$_\beta$ lines
is within $\sim 50$ \kms\ and the two measurements
are consistent within errors (the corresponding channel widths were $\sim 80$ \kms\
for both observations).
The lines are reported at redshifts
$z$({\rm C}\,{\sc ii})$ = 3.0993(4)$, and $z({\rm H}_\beta) = 3.1000(3)$,
and thus their position errors are 30 \kms\ and 23 \kms, respectively.
If we consider the velocity offset of 50 \kms\
as caused by the Doppler noise (see Sect. 3), 
then the limit on \daa\ is defined by the
line position errors which give
$|\Delta\alpha/\alpha| < 6\times10^{-5}$.

Other high-redshift objects, where
both hydrogen and [C\,{\sc ii}] emission lines were detected,
are the galaxies Ly$\alpha$-1 and Ly$\alpha$-2, 
and a sub-millimeter galaxy~-- companions of the quasar BR~1202--0725 at $z = 4.7$ 
(Wagg \etal\ 2012; Carilli \etal\ 2013;  Williams \etal\ 2014). 
The narrowest [C\,{\sc ii}]  158$\mu$m profile  with an FWHM $\sim 56$ \kms\ 
was observed towards the Ly$\alpha$-1 object, yielding a redshift $z = 4.6950(3)$. 
The comparison of the hydrogen Ly$\alpha$ profile with that of [C\,{\sc ii}] 158$\mu$m
shows that their peak positions are slightly shifted 
with respect to each other by $\Delta V = 49$ \kms. 
Again, assuming that this shift is due to the Doppler noise and that the uncertainty of 
$\Delta V$ is of the same order as in the previous case, one obtains 
$|\Delta\alpha/\alpha| \la 6\times10^{-5}$.

\section{Discussion and conclusion}

The application of the FST method based on different species
may yield a biased estimate of \daa\ if the velocity distribution of these species
in the same galaxy is different.
Random Doppler shifts of global velocity profiles (integrated spectra)
caused by non-identical spatial distributions of tracers
can mimic non-zero signals in \daa.
This so-called Doppler noise (e.g., Levshakov \etal\ 2008)
may become a problem when the accuracy of the line position
measurements is increased.

In local disk galaxies the
H\,{\sc i} disk, traced by the hydrogen 21 cm emission,
typically extends significantly beyond the main stellar disk.
On the other hand, the ionized carbon [C\,{\sc ii}] is observed throughout the ISM and
the [C\,{\sc ii}] emission is usually enhanced at the edges of molecular clouds
in the photodissociation regions.
C\,{\sc ii} can be excited by collisions with electrons, 
neutral H\,{\sc i} atoms and molecular H$_2$ and
because of the low ionization potential
of carbon (11.26 eV versus 13.60 eV for hydrogen),
the [C\,{\sc ii}] 158 $\mu$m 
line arises both from ionized and neutral gas,
thus tracing most of the phases of the ionized and atomic ISM 
(e.g., Pineda \etal\  2013; Kaufman \etal\ 1999).
But these are the regions of enhanced hydrogen Balmer lines as well.
A co-spatial distribution of the [C\,{\sc ii}] emission 
with regions seen in H$_\alpha$ emission
are known to be observed in some dwarf galaxies (e.g., Cigan \etal\ 2016).
This tendency might be expected since both the H$_\alpha$ and [C\,{\sc ii}] emission
is generally confined to the optical disks of galaxies (e.g., de Blok \etal\ 2016).
Thus, we may suppose a close correlation between the [C\,{\sc ii}] 158 $\mu$m
and H$_\beta$ spatial distributions over the surface of distant galaxies as well.
We can also expect that
the value of the Doppler noise for the global velocity profiles of the
[C\,{\sc ii}] 158$\mu$m and H$_\beta$ emission lines does not exceed a few \kms\, which is less
than the current error of the line position measurement, $\sigma_{\rm v} \sim 20-30$ \kms\
(see Sec.~2).

As for the hydrogen Ly$\alpha$ line,
it should be noted that at high redshifts its blue wing 
may suffer from some intergalactic medium attenuation
which causes the Ly$\alpha$ centroid to be artificially redshifted.
Besides, additional
uncertainties in interpreting differences among profiles may appear if
different components of distant sources had different values of internal dust
extinction.
Velocity offsets between emission lines 
relative to Ly$\alpha$ were found by several authors
(e.g., Stark et al. 2017; Erb et al. 2014).
From this point of view the mid-infrared 
and far-infrared atomic transitions are
more preferable as reference lines in probing \daa.
In any case to quantify uncertainties induced by the Doppler noise a sample
of \daa\ measurements is needed.

To conclude we note that the current spectral observations of
the dusty star-forming galaxy LAB1-ALMA3 at $z = 3.1$,
and the Ly$\alpha$-1 galaxy at $z = 4.7$  reveal no evidence or a variability
of $\alpha$ and constrain the value of \daa\ 
at the level of $6\times10^{-5}$.

\section*{Acknowledgements}
We thank our referee Christian Henkel
for valuable comments and suggestions that improved the paper.


\begin{thebibliography}{}

\bibitem{}
Agafonova I. I., Molaro P., Levshakov S. A., Hou J. L., 2011, A\&A, 529, 28

\bibitem{}
Aravena M., Decarli R., Walter F., \etal, 2016, ApJ, 833, 153

\bibitem{}
Bahcall J. N., Steinhardt C. L., Schlegel D., 2004, ApJ, 600, 520

\bibitem{}
Bahcall J. N., Sargent W. L. W.,  Schmidt M., 1967, ApJ, 149, L11

\bibitem{}
Bahcall J. N.,  Schmidt M., 1967, PhRvL, 19, 1294

\bibitem{}
Bahcall J. N.,  Salpeter E. E., 1965, ApJ, 142, 1677

\bibitem{}
Bonifacio P., Rahmani H., Whitmore J. B., \etal, 2014, Astron. Nachrichten, 335, 83

\bibitem{}
Braine J., Gratier P.,  Kramer C., \etal, 2012, A\&A, 544, A55

\bibitem{}
Capak P. L., Carilli C., Jones G., \etal, 2015, Nature, 522, 455

\bibitem{}
Carilli C. L., Riechers D., Walter F., \etal, 2013, ApJ, 763, 120

\bibitem{}
Carniani S., Maiolino R., Pallottini A., \etal, 2017, arXiv:1701.03468

\bibitem{}
Cigan P., Young L., Cormier D., \etal, 2016, AJ, 151, 14

\bibitem{}
Cooksy A. L., Blake G. A., Saykally R. J., 1986, ApJL, 305, L89

\bibitem{}
Davis E. D., Hamdan L., 2015, PhRvC, 92, 014319

\bibitem{}
de Blok W. J. G., Walter F., Smith J.-D. T., \etal, 2016, AJ, 152, 51

\bibitem{}
Dzuba V. A., Flambaum V. V., Webb, J. K., 1999a, PhRvA, 59, 230

\bibitem{}
Dzuba V. A., Flambaum V. V., Webb, J. K., 1999b, PhRvL, 82, 888

\bibitem{}
Erb D. K., Steidel C. C., Trainor R. F., \etal, 2014, ApJ, 795, 33

\bibitem{}
Evans T., M., Murphy M. T., Whitmore J. B., \etal, 2014, MNRAS, 445, 128

\bibitem{}
Gao H., Wu X. F.,  M\'esz\'aros P., 2015, ApJ, 810, 121

\bibitem{}
Geach J. E., Narayanan D., Matsuda Y., \etal, 2016, ApJ, 832, 37

\bibitem{}
Godun R. M., Nisbet-Jones P. B. R., Jones J. M., \etal, 2014, PhRvL, 113, 210801

\bibitem{}
Huang Y., Guan H., Liu P., \etal, 2016, PhRvL, 116, 013001

\bibitem{}
Huntemann N., Sanner C., Lipphardt B., Tamm C., Peik E., 2016, PhRvL, 116, 063001

\bibitem{}
Iono D., Yun M. S., Elvis M., \etal, 2006, ApJ, 645, L97

\bibitem{}
Kaufman M. J.,Wolfire M. G., Hollenbach D. J., Luhman, M. C., 1999, ApJ, 527, 795

\bibitem{}
Knudsen K. K., Watson D., Frayer D., \etal, 2017, MNRAS, 466, 138

\bibitem{}
Knudsen K. K., Richard J., Kneib J.-P., \etal, 2016, MNRAS, 462, L6

\bibitem{}
Kosteleck\'y V. A.,  Lehnert  R., Perry M. J., 2003, PhRvD, 68, 123511

\bibitem{}
Kozlov M. G., Porsev S. G., Levshakov S. A., \etal, 2008, PhRvA, 77, 032119

\bibitem{}
Kubo M., Yamada T., Ichikawa T., \etal, 2015, ApJ, 799, 38

\bibitem{}
Leite A. C. O., Martins C. J. A. P., Molaro P., Corre D., Cristiani S., 2016, arXiv:1612.05284

\bibitem{}
Levshakov S. A., Combes F., Boone F. \etal, 2012, A\&A, 540, L9

\bibitem{}
Levshakov S. A., Reimers D., Kozlov M. G., \etal, 2008, A\&A, 479, 719

\bibitem{}
Levshakov S. A., Centuri\'on M., Molaro P., \etal, 2006, A\&A, 449, 879

\bibitem{}
Levshakov S. A. 1994, MNRAS, 269, 339

\bibitem{}
Levshakov, S. A., 1992, in High Resolution Spectroscopy with the VLT, ed.
M.-H. Ulrich, ESO: Garching/Munchen, p. 139

\bibitem{}
Liberati S., 2013, Class. Quant. Grav., 30, 133001

\bibitem{}
Maiolino R., Carniani S., Fontana A., \etal, 2015, MNRAS, 452, 54

\bibitem{}
Maiolino R., Cox P., Caselli P., \etal, 2005, A\&A, 440, L51

\bibitem{}
Molaro P., Centuri\'on M., Whitmore J. B., \etal, 2013, A\&A, 555, A68

\bibitem{}
Mookerjea B., Israel F., Kramer C., \etal, 2016, A\&A, 586, A37

\bibitem{}
Nemitz N., Ohkubo T., Takamoto M., \etal, 2016, Nat. Photonics, 10, 258

\bibitem{}
Nicholson T. L., Campbell S. L., Hutson R. B., \etal, 2015, Nat. Comm., 6, 6896

\bibitem{}
Nisbet-Jones P. B. R., King S. A., Jones J. M., \etal, 2016, Applied Phys. B, 122, 57

\bibitem{}
Pavesi R., Riechers D. A., Capak P. L., \etal, 2016, ApJ, 832, 151

\bibitem{}
Petitjean P., Wang F. Y.,  Wu X. F., Wei J. J., 2016, SpScRv, 202, 195

\bibitem{}
Pineda J. L., Langer W. D., Velusamy T., Goldsmith P. F.,  2013, A\&A, 554, A103

\bibitem{}
Quast R., Reimers D., Levshakov S. A., 2004, A\&A, 415, L7

\bibitem{}
Riechers D. A., Carilli C. L., Capak P. L., \etal, 2014, ApJ, 796, 84

\bibitem{}
Rosenband T., Hume D. B., Schmidt P. O., \etal, 2008, Science, 319, 1808

\bibitem{}
Savedoff M. P., 1956, Nature, 176, 688

\bibitem{}
Sobelman I. I., 1979, Atomic spectra and radiative transitions, Springer-Verlag, Berlin

\bibitem{}
Stark D. P., Ellis R. S., Charlot S., \etal, 2017, MNRAS, 464, 469

\bibitem{}
Tyumenev R., Favier M., Bilicki S., \etal, 2016, NJPh, 18, 113002

\bibitem{}
Umehata H., Matsuda Y., Tamura Y., \etal, 2017, ApJ, 834, L16

\bibitem{}
Venemans B. P., McMahon R. G., Walter F., \etal, 2012, ApJL, 751, L25

\bibitem{}
Wagg J, Wiklind T., Carilli C. L., \etal, 2012, ApJ, 752, L30

\bibitem{}
Wei J. J., Wang J. S., Gao H., Wu X. F., 2016, ApJ, 818, L2

\bibitem{}
Wei J. J., Gao H., Wu X. F.,  M\'esz\'aros P., 2015, PhRvL, 115, 261101

\bibitem{}
Wei\ss\ A., Walter F., Downes D., \etal, 2012, ApJ, 753, 102

\bibitem{}
Whitmore J. B., Murphy, M. T., 2015, MNRAS, 447, 446

\bibitem{}
Williams R. J., Wagg J., Maiolino R., \etal, 2014, MNRAS, 439, 2096

\bibitem{}
Willott C. J., Carilli C. L., Wagg J.,  Wang R., 2015, ApJ, 807, 180

\end{thebibliography}
\end{document}